\documentclass[letterpaper,10pt,twocolumn,twoside]{IEEEtran}

\usepackage{epsfig,amssymb,psfrag,pstricks,cite}

\newtheorem{theorem}{Theorem}
\newtheorem{lemma}{Lemma}
\def\ts{\textstyle}
\def\beq{\begin{equation}}
\def\eeq{\end{equation}}
\def\beqa{\begin{eqnarray}}
\def\eeqa{\end{eqnarray}}
\def\beqan{\begin{eqnarray*}}
\def\eeqan{\end{eqnarray*}}

\newcommand{\simlabel}[1]{ \stackrel{(#1)}{\sim} }
\def\half{{\ts\frac{1}{2}}}

\renewcommand{\P}[1]{\textbf{P}\left({#1}\right)} 
\newcommand{\E}[1]{\textbf{E}\left[{#1}\right]}   

\def\Z{\mathbb{Z}}

\begin{document}

\title{Scalar Quantization with Random Thresholds}

\author{Vivek K Goyal
  \thanks{This material is based upon work supported by the
    National Science Foundation under Grant No.\ 0729069.}
  \thanks{V. K. Goyal is with the
    Massachusetts Institute of Technology
    (e-mail: vgoyal@mit.edu).}
}

\maketitle

\begin{abstract}
The distortion--rate performance of certain
randomly-designed scalar quantizers is determined.
The central results are the mean-squared error distortion and
output entropy for quantizing a uniform random variable with
thresholds drawn independently from a uniform distribution.
The distortion is at most 6 times that of an
optimal (deterministically-designed) quantizer, and
for a large number of levels the output entropy is
reduced by approximately $(1-\gamma)/(\ln 2)$ bits,
where $\gamma$ is the Euler--Mascheroni constant.
This shows that the high-rate asymptotic distortion
of these quantizers in an entropy-constrained context
is worse than the optimal 
quantizer by at most a factor of $6e^{-2(1-\gamma)} \approx 2.58$.
\end{abstract}


\section{Introduction}
What is the performance of a collection of $K$ subtractively-dithered
uniform scalar quantizers with the same step size, used in parallel?
The essence of this question---and a precise analysis under
high-resolution assumptions---is captured by
answering another fundamental question:
What is the mean-squared error (MSE) performance
of a $K$-cell quantizer with randomly-placed thresholds
applied to a uniformly-distributed source?
For both (equivalent) questions,
it is not obvious \emph{a~priori} that the performance penalties
relative to optimal deterministic designs are bounded;
here we find concise answers that demonstrate that these
performance penalties are small.
Specifically, the multiplicative penalty in MSE
for quantization of a uniform source
is at most $6$ in the codebook-constrained case
and about $6e^{-2(1-\gamma)} \approx 2.58$ in the
entropy-constrained case at high rate,
where $\gamma$ is the Euler--Mascheroni constant~\cite{Havil:03}.
The translation of these results is that the multiplicative penalty in MSE
for high-rate parallel dithered quantization
is at most $6$ when there is no expoitation
of statistical dependencies between channels
and about $6e^{-2(1-\gamma)}$ when joint entropy coding or
Slepian--Wolf coding~\cite{SlepianW:73}
is employed and the number of channels is large.

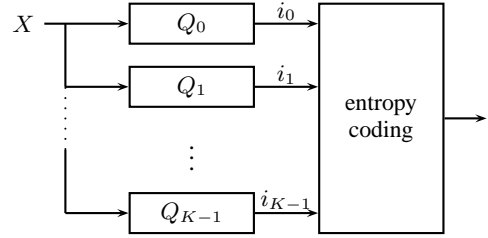
\begin{figure}
  \begin{center}
    \psset{unit=5.6mm}
    \begin{pspicture}(-2.5,-1)(8.5,4.5)
      \rput(-2.5,4.0){\small $X$}

      \psline{->}(-2.0,4.0)(0.0,4.0)
      \psframe(0,3.5)(3,4.5)
      \rput(1.5,4.0){\small $Q_0$}
      \psline{->}(3.0,4.0)(4.5,4.0)
      \rput(3.75,4.3){\small $i_0$}

      \psline{-}(-1.5,4.0)(-1.5,2.5)
      \psline{->}(-1.5,2.5)(0.0,2.5)
      \psframe(0,2)(3,3)
      \rput(1.5,2.5){\small $Q_1$}
      \psline{->}(3.0,2.5)(4.5,2.5)
      \rput(3.75,2.8){\small $i_1$}

      \rput(1.5,1.0){$\vdots$}

      \psline[linestyle=dotted]{-}(-1.5,2.5)(-1.5,1.0)
      \psline{-}(-1.5,1.0)(-1.5,-0.5)
      \psline{->}(-1.5,-0.5)(0.0,-0.5)
      \psframe(0,-1)(3,0)
      \rput(1.5,-0.5){\small $Q_{K-1}$}
      \psline{->}(3.0,-0.5)(4.5,-0.5)
      \rput(3.75,-0.2){\small $i_{K-1}$}

      \psframe(4.5,-1)(7.5,4.5)
      \rput(6.0,2.05){\small entropy}
      \rput(6.0,1.45){\small coding}
      \psline{->}(7.5,1.75)(8.5,1.75)
    \end{pspicture}
  \end{center}
  \caption{Use of $K$ dithered uniform scalar quantizers in parallel.
    Quantizer $Q_k$ has thresholds $\{(j+a_k)\Delta\}_{j \in \Z}$,
    with $a_k$ its \emph{offset}.}
  \label{fig:parallelQuantizers}
\end{figure}

Quantization with parallel channels
is illustrated in Fig.~\ref{fig:parallelQuantizers}.
Each of $K$ quantizers is a subtractively-dithered uniform scalar
quantizer with step size $\Delta$.
Denoting the dither, or \emph{offset}, of quantizer $Q_k$ by $a_k$,
the thresholds of the quantizer are $\{(j+a_k)\Delta\}_{j \in \Z}$.
One may imagine several stylized applications in which it is advantageous
to allow the $a_k$s to be arbitrary or chosen uniformly at random.
For example, with parallel quantizer channels,
one may turn channels on and off adaptively based on available
power or the desired signal fidelity~\cite{ChandrakasanSB:92}.
Alteration of the lossless coding block could then be achieved
through a variety of means~\cite{ZivL:77,ClearyW:84,WittenNC:91}.
The same figure could represent a distributed setting, in which $K$
sensors measure highly-correlated quantities (all modeled as $X$);
with a Slepian--Wolf code~\cite{SlepianW:73} or universal
Slepian--Wolf code~\cite{OohamaH:94}, the sensors can quantize and encode
their samples autonomously.
Variations in the $a_k$s could also arise unintentionally, through
process variation in sensor manufacturing due to cost reduction or
size reduction; mitigation of process variations is expected to be of
increasing importance~\cite{ITRS2010}.
This letter addresses the performance loss relative to deterministic
joint design of the channels or coordinated action by the distributed sensors.

Collectively, the $K$ parallel quantizers specify input $X$ with
thresholds $\cup_{k=0}^{K-1} \{(j+a_k)\Delta\}_{j \in \Z}$.
One would expect the best performance from having
$\{a_k\}_{k=0}^{K-1}$ uniformly spaced in $[0,1]$ through $a_k = k/K$;
this intuition is verified under high-resolution assumptions,
where the optimal entropy-constrained quantizers are uniform~\cite{GishP:68}.
To analyze performance relative to this ideal,
it suffices to study one interval of length $\Delta$ in the domain of
the quantizers because the thresholds repeat with a period of $\Delta$.
This analysis is completed in Section~\ref{sec:basic}.
The ramifications for
the system in Fig.~\ref{fig:parallelQuantizers}
are made explicit
in Section~\ref{sec:DitheredQuantizers}.
Section~\ref{sec:twoUnequal} considers uniform quantizers with unequal step
sizes, and Section~\ref{sec:Discussion} provides additional connections
to related results and concludes the note.

\section{Random Quantizer for a Uniform Source}
\label{sec:basic}

Let $X$ be uniformly distributed on $[0,1)$.
Suppose that a $K$-level quantizer for $X$ is designed by choosing
$K-1$ thresholds independently, each with a uniform
distribution on $[0,1)$. 
Put in ascending order, the random thresholds are denoted
$\{a_k\}_{k=1}^{K-1}$, and
for notational convenience, let $a_0 = 0$ and
$a_K = 1$.
A regular quantizer with these thresholds has lossy encoder
$\alpha: [0,1) \rightarrow \{1,\,2,\,\ldots,\,K\}$ given by
$$
  \alpha(x) \ = \ k
\qquad
  \mbox{for $x \in [a_{k-1},a_k)$}.
$$
The optimal reproduction decoder for MSE distortion is
$\beta: \{1,\,2,\,\ldots,\,K\} \rightarrow [0,1)$ given by
$$
  \beta(k) \ = \ \half(a_{k-1}+a_k).
$$
We are interested in the average rate and distortion of this random quantizer
as a function of $K$, both with and without entropy coding.

\begin{theorem}
\label{thm:basicD}
The MSE distortion, averaging over both the source
variable $X$ and the quantizer thresholds $\{a_k\}_{k=1}^{K-1}$, is
\begin{equation}
  \label{eq:theorem-D}
  D \ = \ \E{(X-\beta(\alpha(X))^2}
    \ = \ \frac{1}{2(K+1)(K+2)}.
\end{equation}
\end{theorem}
\begin{IEEEproof}
Let $L(x \, | \, \{a_k\}_{k=1}^{K-1})$ denote the length of the quantizer
partition cell that contains $x$ when the random thresholds are
$\{a_k\}_{k=1}^{K-1})$;
i.e.,
$$L(x \, | \, \{a_k\}_{k=1}^{K-1}) \ = \ {\rm length}(\alpha^{-1}(\alpha(x))).$$
Since $X$ is uniformly distributed and the thresholds are independent of $X$,
the quantization error is conditionally uniformly distributed for any values
of the thresholds.  Thus the conditional MSE given the thresholds is
$\E{L^2 \, | \{a_k\}_{k=1}^{K-1}}/12$,
and averaging over the thresholds as well gives
$D = \E{L^2}/12$.

The possible values of the interval length, $\{a_i - a_{i-1}\}_{i=1}^K$,
are called \emph{spacings} in the
order statistics literature~\cite[Sect.~6.4]{DavidN:03}.
With a uniform parent distribution, the spacings are identically distributed.
Thus they have the distribution of the minimum, $a_1$:
$$
  f_{a_1}(a) \ = \ (K-1)(1-a)^{K-2},
\qquad
  0 \leq a \leq 1.
$$
The density of $L$ is obtained from the density of $a_1$ by noting that
the probability that $X$ falls in an interval is proportional to the
length of the interval:
$$
  f_L(\ell) \ = \ \frac{ \ell f_{a_1}(\ell) }
                       { \int_0^1 \ell f_{a_1}(\ell) \, d\ell }
            \ = \ K(K-1) \ell (1-\ell)^{K-2},
$$
for $0 \leq \ell \leq 1$.
Now
\beqan
  D & = & \frac{1}{12}\E{L^2}
    \ = \ \frac{1}{12} \int_0^1 \ell^2 \cdot K(K-1) \ell (1-\ell)^{K-2} \, d\ell \\
    & = & \frac{1}{12} \cdot \frac{6}{(K+1)(K+2)},
\eeqan
completing the proof.
An alternative proof is outlined in the Appendix.
\end{IEEEproof}

The natural comparison for (\ref{eq:theorem-D}) is against an
optimal $K$-level quantizer for the uniform source.
The optimal quantizer has evenly-spaced thresholds, resulting in partition
cells of length $1/K$ and thus MSE distortion of $1/(12K^2)$.
Asymptotically in $K$,
Distortion (\ref{eq:theorem-D}) is worse by a factor of $6K^2/((K+1)(K+2))$,
which is at most $6$ and approaches $6$ as $K \rightarrow \infty$.
In other words, designing a \emph{codebook-constrained} or
\emph{fixed-rate} quantizer by choosing the thresholds at random
creates a multiplicative distortion penalty of at most 6\@.

Now consider the \emph{entropy-constrained} or \emph{variable-rate} case.
If an entropy code for the indexes is designed without knowing the realization
of the thresholds, the rate remains $\log_2 K$ bits per sample.
However, conditioned on knowing the thresholds, the quantizer index
$\alpha(X)$ is not uniformly distributed, so the performance
penalty can be reduced.

\begin{theorem}
\label{thm:basicR}
The expected quantizer index conditional entropy,
averaging over the quantizer thresholds $\{a_k\}_{k=1}^{K-1}$, is
\begin{equation}
  \label{eq:theorem-R}
  R \ = \ \E{ H\left( \alpha(X) \, | \, \{a_k\}_{k=1}^{K-1} \right) }
    \ = \ \frac{1}{\ln 2} \sum_{k=2}^K \frac{1}{k}.
\end{equation}
\end{theorem}
\begin{IEEEproof}
The desired expected conditional entropy is the expectation of the self-information,
$-\log_2 \P{\alpha(X)}$.
Let $L$ be defined as in the proof of Theorem~\ref{thm:basicD} to be
the length of the interval containing $X$.
Since the probability of $X$ falling into any subinterval of $[0,1)$
of length $c$ is $c$,
we have
\beqan
  R & = & \E{-\log_2 L} \\
    & = & - \int_0^1 (\log_2 \ell) \, \cdot \, K(K-1) \ell (1-\ell)^{K-2} \, d\ell,
\eeqan
which equals (\ref{eq:theorem-R}) by direct calculation;
see also~\cite{EbrahimiSZ:04},\cite[$\S$4.6]{Varshney:06}.
An alternative proof is outlined in the Appendix.
\end{IEEEproof}

To compare again against an optimal $K$-level quantizer,
note that evenly-spaced thresholds would yield $R = \log_2 K$
while the rate in (\ref{eq:theorem-R}) is also essentially logarithmic in $K$.
The quantity (\ref{eq:theorem-R}) includes the
\emph{harmonic number} $H_n = \sum_{k=1}^n 1/k$,
which has been studied extensively.
For example,
$$
  \frac{1}{24(n+1)^2} \ \leq \ H_n - \gamma - \ln(n+\half)
     \ \leq \ \frac{1}{24n^2}
$$
where $\gamma \approx 0.577216$ is called the Euler--Mascheroni
constant~\cite{Havil:03}.

\begin{figure}
 \begin{center}
  \psfrag{a}[bl][bl]{\small $6e^{-2(1-\gamma)}$}
  \epsfig{figure=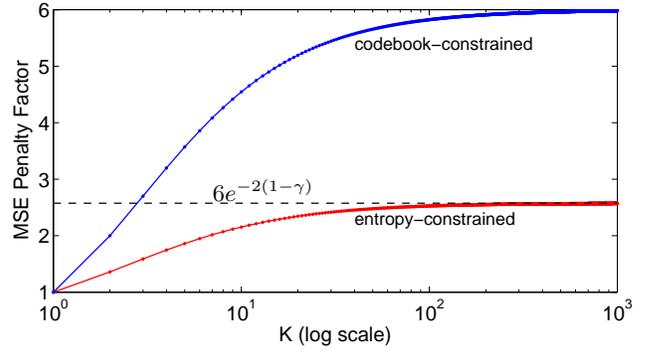,width=3.3in}
 \end{center}
 \caption{MSE penalty factor as a function of $K$.
  For quantization of uniform source on $[0,1]$, $K$ is the number of codewords
  (Section~\ref{sec:basic}).
  For parallel dithered quantization, $K$ is the number of channels
  (Section~\ref{sec:DitheredQuantizers}).}
 \label{fig:randomQuantTheory}
\end{figure}

Combining (\ref{eq:theorem-D}) and (\ref{eq:theorem-R}) while
exploiting the asymptotic approximation $H_n \asymp \gamma + \ln (n+\half)$
yields
$$
  R \ \sim \ (\gamma - 1 + \ln (K+\half))/(\ln 2)
$$
and a distortion--rate performance of
\begin{equation}
  \label{eq:average-highrate-DR}
  D \ \sim \ \half e^{-2(1-\gamma)}2^{-2R},
\end{equation}
where $\sim$ represents a ratio approaching 1 as $K$ increases for distortions
and difference approaching 0 as $K$ increases for rates.
The exact performance from (\ref{eq:theorem-D})--(\ref{eq:theorem-R})
is shown in Fig.~\ref{fig:randomQuantTheory}
with normalization through division by $\frac{1}{12} 2^{-2R}$.

\section{Parallel Dithered Quantizers}
\label{sec:DitheredQuantizers}
Let us now return to the system depicted in Fig.~\ref{fig:parallelQuantizers}.
High-resolution analysis of this system for any number of channels $K$
follows easily from the results of the previous section.

For notational convenience, let us assume that the source $X$
has a continuous density supported on $[0,1)$.
Fix $\Delta \ll 1$ and consider $K$ uniform quantizers with step size
$\Delta$ applied to $X$.
Quantizer $Q_0$ has lossy encoder $\alpha_0$ with thresholds at integer multiples of $\Delta$.
The remaining $K-1$ quantizers are offset by $a_k\Delta$,
i.e., the thresholds of Quantizer $Q_k$ with lossy encoder $\alpha_k$ are at
$\{(j + a_k)\Delta\}_{j=0,1,\,\ldots,\lfloor\Delta^{-1}\rfloor}$.

We would like to first approximate the distortion in
joint reconstruction from
$(\alpha_0(X),\alpha_1(X),\ldots,\alpha_{K-1}(X))$.
The first quantizer index $\alpha_0(X)$ isolates $X$ to an interval
$\alpha_0^{-1}(\alpha_0(X))$ of length $\Delta$.
Since $X$ has a continuous density and $\Delta \ll 1$,
we may approximate $X$ as conditionally uniformly distributed on this interval.
Thus we may apply Theorem~\ref{thm:basicD} to obtain
\begin{equation}
  \label{eq:ditheredEqual-D}
  D \ \sim \ \frac{\Delta^2}{2(K+1)(K+2)},
\end{equation}
where $\sim$ represents a ratio approaching 1 as $\Delta \rightarrow 0$.
The average of the joint entropy is increased from (\ref{eq:theorem-R})
by precisely $H(\alpha_0(X))$.
Since
$$
 \lim_{\Delta \rightarrow 0} H(\alpha_0(X)) - h(X) - \log_2 \Delta^{-1} \ = \ 0,
$$
where $h(X)$ is the differential entropy of $X$~\cite{Renyi:59},
\begin{equation}
  \label{eq:ditheredEqual-R}
  R \ \sim \ h(X) + \log_2 \Delta^{-1} + \frac{1}{\ln 2} \sum_{i=2}^K \frac{1}{i},
\end{equation}
where $\sim$ represents a difference approaching 0 as $\Delta \rightarrow 0$.
For a large number of channels $K$, eliminating $\Delta$ gives
\begin{eqnarray}
  D & \simlabel{a} & \frac{\exp(2\sum_{i=2}^K i^{-1})}{2(K+1)(K+2)} 2^{2h(X)}2^{-2R}
  \label{eq:ditheredEqual_a} \\
    & \simlabel{b} & \frac{\exp(2(\gamma-1+\ln (K+\half)))}{2(K+1)(K+2)} 2^{2h(X)} 2^{-2R} \nonumber \\
    &   =     & \frac{\exp(2(\gamma-1))(K+\half)^2}{2(K+1)(K+2)} 2^{2h(X)} 2^{-2R} \nonumber \\
    & \simlabel{c} & \half e^{-2(1-\gamma)} 2^{2h(X)} 2^{-2R} \nonumber
\end{eqnarray}
where (a) is exact as $\Delta \rightarrow 0$,
(b) is the standard approximation for harmonic numbers, and
(c) is an approximation for large $K$.
This distortion exceeds the distortion of optimal entropy-constrained
quantization by the factor $6e^{-2(1-\gamma)}$.

\section{Quantizers with Unequal Step Sizes}
\label{sec:twoUnequal}
The methodology introduced here can be extended to cases with unequal
quantizer step sizes.
The details become quickly more complicated as the number of
distinct step sizes is increased, so we consider only two step sizes.
We also limit attention to source $X$ uniformly distributed on $[0,1)$.

Let quantizer $\alpha_0$ be a uniform quantizer with step size $\Delta_0 \ll 1$
and thresholds at integer multiples of $\Delta_0$ (no offset).
Let $\alpha_1$ be a uniform quantizer with step size $\Delta_1 \ll 1$ and
thresholds offset by $a_1$, where $a_1$ is uniformly distributed on
$[0,\Delta_1)$.
Without loss of generality, assume $\Delta_0 < \Delta_1$.
(It does not matter which quantizer is fixed to have no offset;
it only simplifies notation.)

Mimicking the analysis in Section~\ref{sec:basic},
the performance of this pair of quantizers
is characterized by the p.d.f.\ of the length of the partition cell
into which $X$ falls.
Furthermore, because of the random dither $a_1$,
the partition cell lengths are identically distributed.

Let $M$ be the length of the partition cell with left edge at $0$.
Clearly $M$ is related to $a_1$ by
\beq
 \label{eq:twoUnequalM}
  M \ = \ \left\{ \begin{array}{ll}
              a_1, & \mbox{if $a_1 \in [0,\Delta_0]$}; \\
         \Delta_0, & \mbox{if $a_1 \in (\Delta_0,\Delta_1)$}.
              \end{array} \right.
\eeq
So $M$ is a mixed random variable with (generalized) p.d.f.\
$$
  f_M(m) \ = \ \frac{1}{\Delta_1} + \left(1-\frac{\Delta_0}{\Delta_1}\right) \delta(m-\Delta_0),
\qquad
  m \in [0,\Delta_0].
$$
With $L$ defined (as before) as the length of the partition cell that contains
$X$,
\beqan
  f_L(\ell) & = & \frac{ \ell f_{M}(\ell) }
                       { \int_0^{\Delta_0} \ell f_{M}(\ell) \, d\ell } \\
            & = & \frac{\ell}{\Delta_0(\Delta_1-\half\Delta_0)}
                + \frac{\Delta_1-\Delta_0}{\Delta_1-\half\Delta_0}
                   \delta(\ell-\Delta_0),
\eeqan
for $0 \leq \ell \leq \Delta_0$.
The average distortion is given by
\begin{equation}
 \label{eq:D-twoQuantizers}
  D \ = \ \frac{1}{12}\E{L^2}
    \ = \ \frac{\Delta_0^2}{12} \, \cdot \,
            \frac{\Delta_1 - \frac{3}{4}\Delta_0}{\Delta_1 - \half\Delta_0}.
\end{equation}
This expression reduces to
(\ref{eq:ditheredEqual-D}) (with $K=2$) for $\Delta_0 = \Delta_1 = \Delta$.
Also, it approaches $\Delta_0^2/12$ as $\Delta_1 \rightarrow \infty$
consistent with the second quantizer providing no information.
The average rate is
\begin{equation}
 \label{eq:R-twoQuantizers}
  R \ = \ \E{-\log_2 L}
    \ = \ \log_2 \Delta_0^{-1}
           + \frac{1}{2\ln2} \, \frac{\Delta_0}{2\Delta_1 - \Delta_0}.
\end{equation}
This reduces to (\ref{eq:ditheredEqual-R})
(with $K=2$ and $h(X) = 1$) for $\Delta_0 = \Delta_1 = \Delta$.

One way in which unequal quantization step sizes could arise is
through the quantization of a frame expansion~\cite{GoyalVT:98}.
Suppose the scalar source $X$ is encoded by dithered uniform
scalar quantization of
$Y = (X \cos \theta,\, X \sin \theta)$
with step size $\Delta \ll 1$ for each component of $Y$.
This is equivalent to using quantizers with step sizes
$$
  \Delta_0 \ = \ \Delta/|\cos\theta|
\qquad
\mbox{and}
\qquad
  \Delta_1 \ = \ \Delta/|\sin\theta|
$$
directly on $X$.
Fixing $\theta \in (0,\pi/4)$ so that $\Delta_0 < \Delta_1$,
we can express the distortion (\ref{eq:D-twoQuantizers})
as
$$
  D_\theta \ = \ \frac{\Delta^2\sec^2\theta}{12} \, \cdot \,
        \frac{1 - \frac{3}{4}\tan\theta}{1 - \half\tan\theta}
$$
and the rate (\ref{eq:R-twoQuantizers}) as
$$
  R_\theta \ = \ \log_2 \Delta^{-1} + \log_2 \cos\theta
           + \frac{1}{2\ln2} \, \frac{\tan\theta}{2 - \tan\theta}.
$$

The quotient
\beq
\label{eq:qTheta}
  q_\theta
           \ = \ \frac{D_\theta}{{\ts \frac{1}{12}}2^{-2R_\theta}}
           \ = \ 
                  \frac{1 - \frac{3}{4}\tan\theta}{1 - \half\tan\theta}
                  \cdot
                  \exp\left(\frac{\tan\theta}{2 - \tan\theta}\right)
\eeq
can be interpreted as the multiplicative distortion penalty as
compared to using a single uniform quantizer.
This is bounded above by
$$
  \left. q_{\theta} \right|_{\theta = \pi/4}
   \ = \ e/2,
$$
which is consistent with evaluating (\ref{eq:ditheredEqual_a}) at $K=2$.
Thus, joint entropy coding of the quantized components largely compensates
for the (generally disadvantageous) expansion of $X$ into a higher-dimensional
space before quantization;
the penalty is only an $e/2$ distortion factor or $\approx 0.221$ bits.

\section{Discussion}
\label{sec:Discussion}
This note has derived distortion--rate performance for
certain randomly-generated quantizers.
The thresholds
(analogous to offsets in a dithered quantizer)
are chosen according to a uniform distribution.
The technique can be readily extended to
other quantizer threshold distributions;
however, the uniform distribution is motivated by the
asymptotic optimality of uniform thresholds in
entropy-constrained quantization.

The analysis in Section~\ref{sec:DitheredQuantizers}
puts significant burden on the
entropy coder to remove the redundancies in the quantizer outputs
$(i_0,\,i_1,\,\ldots,\,i_{K-1})$.
This is similar in spirit to the universal coding scheme of Ziv~\cite{Ziv:85},
which employs a dithered uniform scalar quantizer along with an
ideal entropy coder to always
perform within 0.754 bits per sample of the
rate--distortion bound.
In the case that the quantizers are distributed, we are analyzing
the common strategy for Wyner--Ziv coding~\cite{WynerZ:76} of
quantizing followed by Slepian--Wolf coding;
we obtain a concrete rate loss upper bound of
$\frac{1}{2}\log_2( 6e^{-2(1-\gamma)}) \approx 0.683$ bits per sample
when the rate is high; this is approached when the number of encoders is large.
With non-subtractive dither,
the randomization of thresholds is unchanged
but the reproduction points are not matched to the thresholds.
Thus, the rate computation is unchanged but distortions are increased.

Use of analog-to-digital converter channels with differing quantization
step sizes was studied in~\cite{MaymonO:10}.
Unlike the present note, this work exploits correlation of a
wide-sense stationary input; however, it is limited by a simple
quantization noise model and estimation by linear, time-invariant (LTI)
filtering.

Exact MSE analysis of quantized overcomplete expansions has proven difficult,
so many papers have focused on only the scaling of distortion with
the redundancy of the
frame~\cite{ThaoV:94b,ThaoV:96,GoyalVT:98,RanganG:01}.
The example in Section~\ref{sec:twoUnequal},
could be extendable to more general frame expansions.

\appendix

The proofs of Theorems~\ref{thm:basicD} and~\ref{thm:basicR}
are indirect in that they introduce the random variable $L$ for
the length of the partition cell containing $X$.
A more direct proof is outlined here.

\begin{lemma}
For fixed thresholds $\{a_k\}_{k=1}^{K-1}$,
$$
  \E{(X-\beta(\alpha(X))^2 \, | \, \{a_k\}_{k=1}^{K-1} }
   \ = \ \sum_{k=1}^K \frac{1}{12}\left( a_k - a_{k-1} \right)^3,
$$
$$
  H\left( \alpha(X) \, | \, \{a_k\}_{k=1}^{K-1} \right)
   \ = \ - \sum_{k=1}^K \left( a_k - a_{k-1} \right)
                        \log_2 \left( a_k - a_{k-1} \right).
$$
\end{lemma}
\begin{IEEEproof}
The quantizer maps interval $[a_{k-1},a_k)$ to $k$
so
$$\P{\alpha(X) = k \, | \, \{a_j\}_{j=1}^{K-1} } = a_k - a_{k-1}.$$
The entropy expression is thus immediate.
The distortion expression follows by expanding the expectation using
the law of total expectation with conditioning on $\alpha(X)$:

\beqan
  \lefteqn{\E{(X-\beta(\alpha(X))^2 \mid \{a_j\}_{j=1}^{K-1} }} \\
   & = & \sum_{k=1}^K
          \underbrace{\E{ (X-\beta(\alpha(X))^2 \mid \alpha(X)=k,\, \{a_j\}_{j=1}^{K-1} }}_{\frac{1}{12}(a_k - a_{k-1})^2} \\
   &   & \qquad \cdot
          \underbrace{\P{ \alpha(X) = k \mid \{a_j\}_{j=1}^{K-1} }}_{(a_k - a_{k-1})}.
\eeqan
~\\[-3.0ex]
\end{IEEEproof}

The theorems are proved by averaging over the joint distribution
of the quantizer thresholds $\{a_i\}_{i=1}^{K-1}$,
which is uniform over the simplex
$0 \leq a_1 \leq a_2 \leq \cdots \leq a_{K-1} \leq 1$.

\section*{Acknowledgments}
The author is thankful to John Sun, Lav Varshney,
and an anonymous reviewer for helpful suggestions.

\end{document}